\documentclass[twocolumn,aps,pra]{revtex4}
\usepackage{epsfig}
\usepackage[english]{babel}
\usepackage{latexsym}
\usepackage{graphics}
\usepackage{subfigure}
\usepackage{graphicx}
\usepackage{dcolumn}
\usepackage{amsmath}
\usepackage{hyperref}
\usepackage{amssymb}
\usepackage{color}

\begin{document}
\title{Strong-field response time and its implications on attosecond measurement}

\author{C. Chen$^{1,\dag}$, J. Y. Che$^{1,\dag}$,  X. J. Xie$^{1}$,  S. Wang$^{1,2}$, G. G. Xin$^{3,\ddag}$, and Y. J. Chen$^{1,*}$}

\date{\today}

\begin{abstract}

To measure and control the electron motion in atoms and molecules by the strong laser field on the attosecond time scale is one of the research frontiers of atomic and molecular photophysics. It involves many new phenomena and processes and raises a series of questions of concepts, theories and methods. Recent studies show that the Coulomb potential can cause the ionization time lag (about 100 attoseconds) between instants of the field maximum and the ionization-rate maximum.
This lag can be understood as the response time of the electronic wave function to the strong-field-induced ionization event.
It has a profound influence on the subsequent ultrafast dynamics of the ionized electron and can significantly change the time-frequency properties of electron trajectory (an important theoretical tool for attosecond measurement). Here, the research progress of response time and its implications on attosecond measurement are briefly introduced.

\end{abstract}

\affiliation{1.College of Physics and Information Technology, Shaan'xi Normal University, Xi'an, China\\
2.School of Physics, Hebei Normal University, Shijiazhuang, China\\
3.School of Physics, Northwest University, Xi'an, China}
\maketitle

\section{Attosecond measurement}
\emph{Introduction}.-Attosecond optics is a subject rising with the rapid development of ultraintense and ultrashort laser pulse technology in recent 20 years. It can realize the real-time measurement and ultrafast control of motion of electrons in atoms and molecules, and provides a new tool for human to understand the micro world. Attosecond optics is closely associated with strong-field physics.
The interaction of intense laser fields with atoms and molecules \cite{Keldysh,Ammosov} produces abundant ultrafast physical phenomena, such as above-threshold ionization (ATI) \cite{Schafer,Yang,Lewenstein,Becker}, high-order harmonic generation (HHG) \cite{McPherson,Huillier,Corkum,Lewenstein2} and non-sequential double ionization (NSDI) \cite{Niikura,Zeidler,Becker2}, etc. These phenomena have been further applied to ultrafast measurements of electron dynamics in atoms and molecules \cite{Krausz,Lepine},
with achieving unprecedented time resolution of attosecond. For example,
scientists have developed attoclock technology for studying tunneling time \cite{Eckle}, high-order harmonic spectroscopy technology for monitoring the trajectories of scattered electrons \cite{Shafir}, and attosecond streaking technology for probing Wigner time delay of electrons \cite{Dahlstrom}.

Although attosecond measurement has made impressive discoveries and achievements, it is still in the primary stage of development \cite{Leone}. 1) Theoretically,  according to the principle of quantum mechanics,  there is no time operator, and the ``measurement" of time is the basic problem of quantum mechanics \cite{Muga,Maquet,Pazourek}. This makes the interpretation of the new attosecond measurement results very complicated even for simple systems, leading to many conceptual and theoretical problems, such as which time information can be obtained, which processes can be manipulated, and so on. One of the basic conceptual problems is whether there is a response time of the electronic wave function  to the photoabsorption event in the interaction of a weak laser field and an atom or a molecule. The experimental measurement of Wigner time delay gives a positive answer to this question, but leads to a series of other conceptual questions \cite{Pazourek,Maquet}, such as whether the response time corresponds to the eigenvalue of a quantum time operator, the relationship between the response time and the energy-time uncertainty principle, whether the absolute value of the response time can be measured, and so on. 2) In terms of method, the current ultrafast measurement technology is not directly achieved on the attosecond time scale. After the end of the interaction between light and matter, the frequency domain (energy domain) information of the system is observed, such as photoelectron spectrum, harmonic radiation spectrum or optical absorption spectrum. After obtaining these information, people use the established classical, semiclassical or quantum theoretical models to retrieve the time information of electron motion, and then realize attosecond-resolved measurement \cite{Eckle,Shafir}. Therefore, the theoretical model plays an important role in the current attosecond measurement scheme. The accuracy of the theoretical model will determine the accuracy of the inversion information \cite{Xie}, which will directly affect the interpretation of the experimental results or the description of the measured process. 3) Currently, attosecond measurement is mainly focused on atoms, and attosecond probing of more widely distributed and more abundant molecules needs to be deepened. Molecules, especially polar molecules \cite{Wang1,Wang2}, have more degrees of freedom, so they will present many new effects in the intense laser field, such as two-center interference \cite{Lein}, alignment and orientation \cite{Chen,Frumker}, permanent dipole (PD) \cite{Etches} and nuclear vibration \cite{Li}. For complex molecules, such as top molecules, they also have different configurations. In the attosecond measurement of molecules, it is necessary to fully consider these effects related to molecules, develop suitable theoretical models to describe, find suitable characteristic quantities to characterize and construct suitable probing schemes to measure these effects. These effects can be used not only as the ``object" of attosecond measurement, but also as an effective means to obtain the structural or dynamical information of the target.

\begin{figure*}[t]
\begin{center}
\rotatebox{0}{\resizebox *{17.5cm}{7cm} {\includegraphics {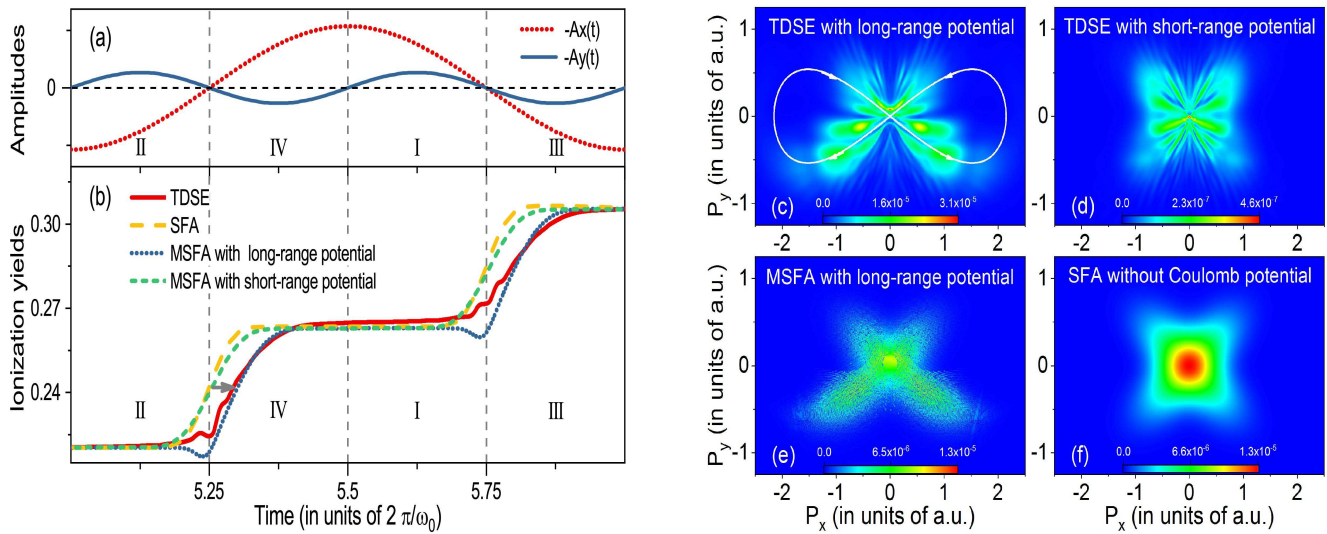}}}
\end{center}
\caption{Coulomb-induced ionization time lag (strong-field response time) and its effects on ATI. \textbf{Left}: (a) Vector potential diagram of the orthogonal two-color laser field used in the calculation. (b) Time-dependent ionization probabilities. It can be seen from (b) that the time of significant increase of ionization yield calculated by TDSE has a time lag relative to the peak time of the laser electric field (5.25T), as shown by the horizontal arrow. The modified SFA with long-range Coulomb potential (long MSFA) well describes the lag, while the modified SFA with short-range potential (short MSFA) or the general SFA without considering the Coulomb effect predicts the strongest ionization at the peak of the laser electric field. \textbf{Right}: PMDs calculated by TDSE with long-range Coulomb potential (c), TDSE with short-range Coulomb potential (d), MSFA with long-range Coulomb potential (e) and SFA (f). Due to the Coulomb-induced ionization time lag, which is associated with the long-range property of the Coulomb potential and increases the contributions of long trajectory, the PMDs in (c) and (e) show a striking up-down asymmetry, which disappears in (d) and (f). (see Ref. \cite{Xie})
}
\label{fig:g1}
\end{figure*}

\emph{Theoretical tools}.-As mentioned above, the theoretical model plays an important role in the current attosecond measurement experiments. Next, we introduce the strong-field theory model and the basic idea of attosecond measurement. At present, the commonly used models include semiclassical models \cite{Schafer,Corkum} and quantum models based on strong-field approximation (SFA) \cite{Lewenstein,Becker}. Using these models, the motion of electrons can be described by the concept of electron trajectories which can be divided into the types of  long trajectory and  short trajectory. These trajectories are characterized by the frequency-domain observable such as photoelectron momentum (harmonic energy) and the ionization time (the ionization time and the return time) of the electron, with establishing a connection between the frequency domain and the time domain of the observing process  \cite{Xie}. By virtue of these electron trajectories, the time information of the system can be retrieved from the observables. The Coulomb effect is not considered in the usual strong-field model, but it has been shown that \cite{Brabec,Milosevic} this effect has an important influence on the photoelectron energy spectrum \cite{Blaga}, photoelectron momentum distribution (PMD) \cite{Goreslavski} and photoelectron angular distribution [16]. In recent years, semiclassical theoretical models for strong-field ionization including Coulomb effect have been developed \cite{Brabec,Goreslavski,yantm2010}. These models play an important role in understanding the ultrafast dynamics of electrons under the action of laser field and Coulomb field. However, previous studies mainly focus on the influence of Coulomb potential on the frequency-domain properties of the electron trajectory, while there are few studies on the time-domain ones. As mentioned above, the time-frequency properties of the electron trajectory are the key to establish the corresponding relationship between observable and time.

\begin{figure*}[t]
\begin{center}
\rotatebox{0}{\resizebox *{18cm}{7cm} {\includegraphics {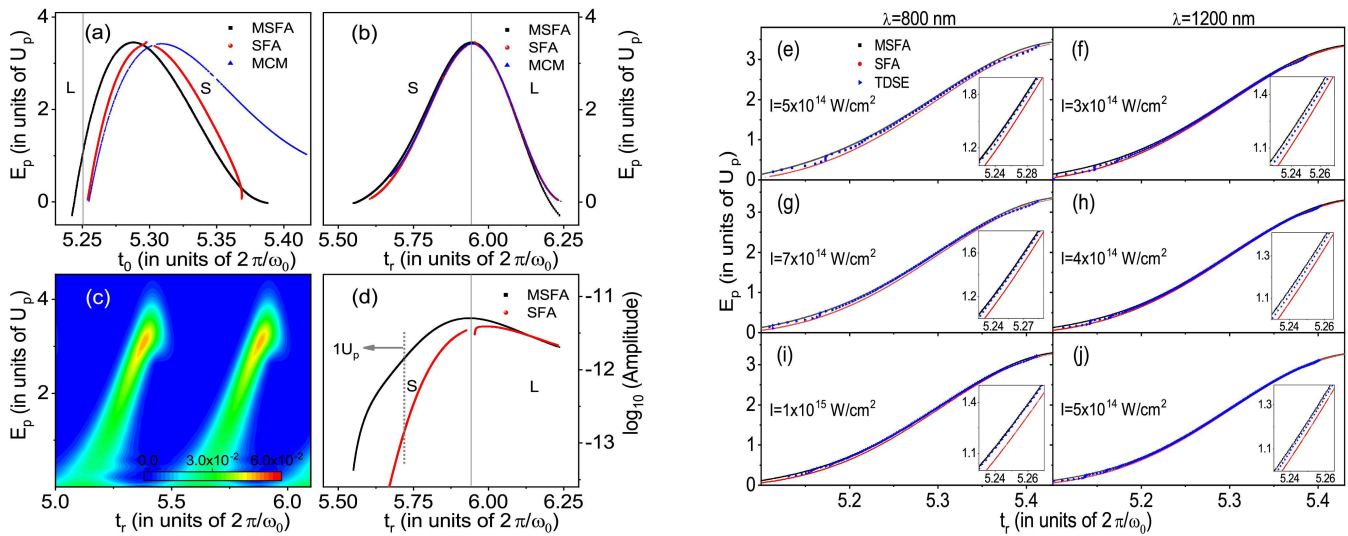}}}
\end{center}
\caption{Effects of strong-field response time on HHG. \textbf{Left}: (a) The HHG return energy versus the tunneling-out time predicted by different models. (b) The HHG return energy versus the return time predicted by different models. (c) The wavelet-analysis result of TDSE dipole acceleration for the short HHG electron trajectory. (d) The HHG return probability versus the return time predicted by different models. Here, MSFA refers to the modified SFA with long-range Coulomb potential, and MCM refers to the semiclassical three-step model with considering the exit position. It can be seen from (a) that the tunneling-out time of the MSFA prediction is more than ten attoseconds earlier than that of the SFA prediction. Correspondingly, in (d), the short-trajectory HHG amplitude of the MSFA prediction is one order of magnitude higher than the SFA one, and is closer to the TDSE result in (c). The HHG return time predicted by different methods is also different. Compared to SFA, the return time predicted by MSFA is also earlier, as shown in (b). In particular, due to the Coulomb-induced ionization time lag, electrons tunneling out of the laser-Coulomb-formed barrier at the rising part of the laser field  can also contribute to the HHG. As seen in (a), the MSFA predicts some long trajectories which begin at times somewhat earlier than 5.25T, while such trajectories are prohibited in SFA and MCM. \textbf{Right}: (e)-(j) Comparisons for short-trajectory HHG return energy versus return time, calculated with MSFA, SFA and TDSE at different laser parameters. The TDSE results are obtained with finding the local maximal amplitudes from the wavelet-analysis results of TDSE dipole acceleration as shown in (c) on the left. They provide a reference for checking the predictions of MSFA. In all cases, the MSFA results are nearer to the TDSE ones than the SFA predictions. Linearly-polarized single-color laser fields are used here. (see Ref. \cite{Xie})
}
\label{fig:g2}
\end{figure*}

\section{Strong-field response time and its effects on attosecond measurement}
\emph{Semiclassical and quantum pictures}.-Recently, through numerical solution of time-dependent Schr\"{o}dinger equation (TDSE) and using modified SFA (MSFA) that includes the Coulomb effect, studies on the influence of coulomb potential on the time-domain properties of
the electron trajectory show that  \cite{Xie}, the Coulomb effect can lead to a significant lag of the ionization time of the system (about 100 attoseconds) relative to the peak time of the laser electric field (Fig. 1). From the semiclassical point of view, the lag means that the
electron does not escape immediately after tunneling out of the laser-Coulomb-formed barrier, but vibrates near the parent nucleus for a period of time under the combined action of Coulomb field and laser electric field,
and then escapes. From the quantum point of view, the ionization time lag can be understood as the response time of the electron wave function to the ionization event
when an atom or a molecule with long-range Coulomb potential interacts with the strong laser field.

This lag is different from the theoretical Wigner time delay, which is only related to the interaction between electrons and Coulomb potential, and essentially belongs to the steady-state problem  \cite{Dahlstrom}.
Compared with the experimental Wigner time delay probed by attosecond streaking technique,
this lag (which can be termed as strong-field response time) exists in the middle of the  strong-laser-induced physical process (such as ATI and HHG),
and has a profound influence on the subsequent ultrafast dynamics of the electron after tunneling,
while the experimental Wigner time delay (weak-field response time) exists at the end of the whole process of bound-continuum transition
triggered by a weak attosecond pulse, then is probed with a weak infrared (IR) laser pulse. More properties of strong-field response time will be discussed later.

\begin{figure*}[t]
\begin{center}
\rotatebox{0}{\resizebox *{17.5cm}{7cm} {\includegraphics {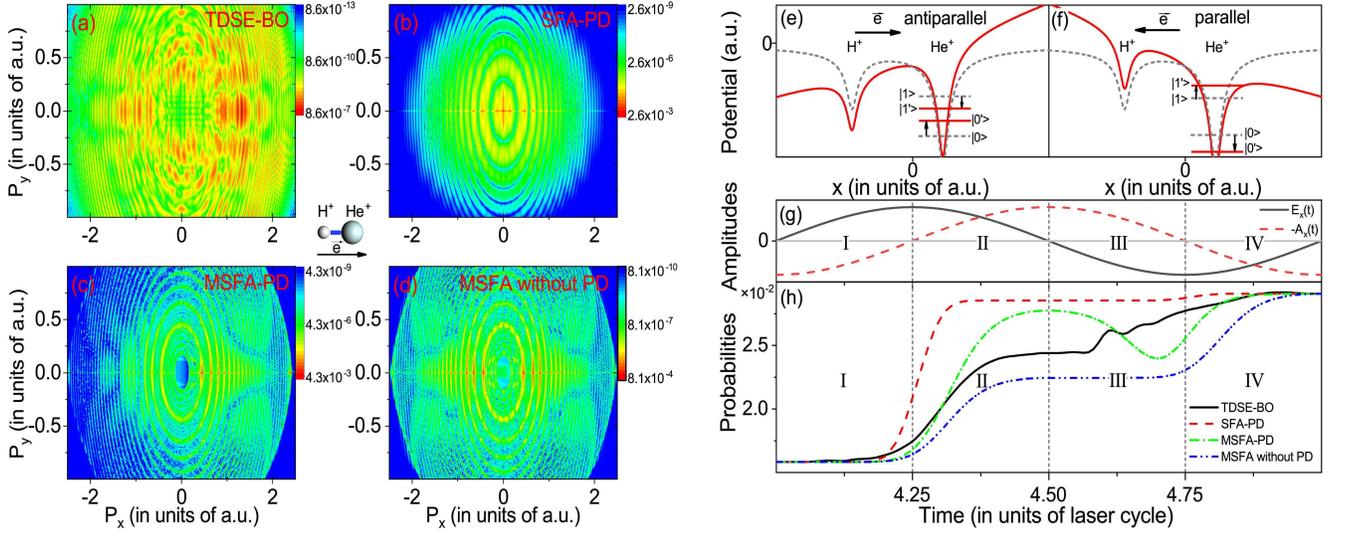}}}
\end{center}
\caption{Effects of strong-field response time on ATI of polar molecules. \textbf{Left}: In (a), the PMD of HeH$^+$ calculated by TDSE under Born-Oppenheimer (BO) approximation shows a significant left-right asymmetry, which is absent in  predictions of SFA-PD and MSFA without PD in (b) and (d). This asymmetry is reproduced in (c) by MSFA-PD including both Coulomb and PD. \textbf{Right}: Due to the PD induced Stark shift, in the first half of the laser period in (e), the ground-state energy is dressed upward and the excited-state energy is dressed downward, and the system is easier to ionize. The situation reverses in the second half cycle in (f). The laser electric field $E_x(t)$ and negative vector potential $-A_x(t)$ used in calculations are shown in (g), assuming that the laser polarization is along the $x$ axis. Classically, the drift momentum of the electron in the $x$ direction satisfies $p_x=-A_x(t)$. The time-dependent ionization probabilities calculated are shown in (h). The TDSE results here also show a significant time lag. In the first (second) half of the period, the ionization probability is large (small) and the time lag is also significant (small). The typical phenomena are only reproduced by MSFA-PD.  In (g) and (h), the region of a laser cycle is divided into four parts (I-IV). In regions I and IV,  $p_x=-A_x(t)<0$, while in regions II and III, $p_x=-A_x(t)>0$. The time-dependent ionization probabilities of TDSE and MSFA-PD show that the amplitudes of photoelectrons with momentum $p_x>0$ (produced in regions II and III) are significantly larger than those with momentum $p_x<0$ (produced in regions I and IV). This indicate that the asymmetry of PMD in (a) on the left is due to the interaction between Coulomb-induced ionization time lag (response time) and PD-induced asymmetric ionization. Linearly-polarized single-color laser fields are used here. (see Ref. \cite{Wang2})
}
\label{fig:g4}
\end{figure*}

\emph{Effects on ATI}.-Next, we discuss the important effect of Coulomb-induced ionization time lag on the subsequent
ultrafast dynamics of the system. This lag can significantly change the time-frequency characteristics of the electron trajectory,
and change a large number of short trajectories (related to the directly ionized electrons) into long ones (related to the ionized electrons with a scattering event).
As a result, the contribution of the long trajectory to the ionization of the system is enhanced by more than 20$\%$,
which significantly changes the structure of the PMD (for example, leading to the significant up-down asymmetry of the PMD in the orthogonal two-color field as shown in  \cite{Xie}, and the significant deflection of the PMD in the elliptical field). In particular,  further studies show that the theoretical prediction of the Coulomb-induced ionization time lag depends on the form of the Coulomb potential used in the theoretical calculation, and different Coulomb potentials will lead to different ionization time lags. This dependence gives  suggestions for  experiments of attosecond measurement based on photoelectron spectroscopy. For example, in the exquisite attoclock experiment \cite{Eckle}, in order to extract the tunneling-time information, people need to subtract the offset angle obtained by semiclassical simulations from the offset angle obtained by experiments (this operation can be understood as eliminating the influence of strong-field response time). A natural problem is which form of Coulomb potential should be selected in semiclassical simulations, and how can one ensure that the influence of response time is completely eliminated. In addition, the orientation, the PD and the vibration of nuclei for molecules may also affect the form of the effective potential, which needs to be fully considered in the future theoretical research. Recent experimental studies also show that the measurement of tunneling time in attosecond experiment is affected by the theoretical model used in inversion \cite{Sainadh,Han}.

\emph{Effects on HHG}.-In addition to strong-field ionization, the Coulomb-induced ionization time lag also has an important influence on the time-domain properties of HHG electron trajectory \cite{Xie}. It leads to more than ten attoseconds advance of the tunneling-out time of the HHG electron trajectory, and makes the tunneling-out time closer to the peak-value time of the external field. As a result, the contribution of the short trajectory (corresponding to the trajectory with a shorter excursion time) to the harmonic radiation is increased by an order of magnitude (Fig. 2).
These effects also give suggestions for experiments of attosecond measurement based on high-order harmonic spectroscopy (HHS). For example, in the beautiful HHS experiment in \cite{Shafir}, the reconstruction scheme of ``rescattering electron trajectory" is designed based on the electron trajectory without Coulomb effect. A natural problem is that if the response time is included in  the reconstruction scheme (if there is no special explanation, the later mentioned response time refers to the Coulomb-induced strong-field ionization time lag), what will happen? It can be expected that there will be attosecond-scale differences between HHG electron trajectories of atoms and molecules with  the same ionization potential but different forms of the Coulomb potential. Future HHS measurement schemes need to be able to distinguish these differences. In addition, how to design an experimental scheme to extract the effect of response time on the HHG yields is also an interesting and challenging work. 
Recent studies based on R-matrix theory that considers the Coulomb effect have also shown that the Coulomb potential induces an important temporal correction to HHG electron trajectory \cite{Torlina}.

\emph{Effects on ATI of polar molecules}.-Finally, recent studies on strong-field ionization of polar molecules show that \cite{Wang2}, the PMD of HeH$^+$ in the linearly-polarized monochromatic laser field presents a significant left-right asymmetry, which is due to the complex interaction between the Coulomb potential and the PD of polar molecules, and the response time is the basis of this typical feature (Fig. 3).  Further studies also show that due to the PD effect, there is a difference of the response time of polar molecules between the first half of the laser cycle and the second half. The difference can be characterized by the ratio of the PMD in different quadrants in the orthogonal two-color field \cite{Che1}, and also by the offset angle of the PMD in the elliptically-polarized laser field \cite{Che2}. Considering the Coulomb-induced temporal correction to electron trajectory, the usual attoclock can be calibrated  to establish a direct one-to-one correspondence between time and photoelectron momentum. Then the response time can be read directly from the Coulomb-corrected attoclock, especially the relative response time between different electronic states \cite{Che2}. 

The response time also plays an important role in strong-field double ionization of HeH$^+$ \cite{Wang3}.  Due to the existence of the response time, when the single ionization of HeH$^+$ is preferred for the first electron escaping along the He side, these two electrons in double ionization prefer to release together along the H side.

Due to the effect of permanent dipole, the dynamics of polar molecules in the intense laser field includes the basic quantum mechanical processes such as tunneling, excitation and Stark shift, and the motion of electrons relative to different nuclei forms a natural contrast. This provides an ideal platform for the study of response time, which is not only convenient to study the internal relationship between these basic processes and response time, but also convenient to study the relative response time, and  to test the applicability of the strong-field theoretical model \cite{Che1,Che2}.

\emph{Properties}.-It should be emphasized that the response time is a relative quantity, which is defined by the comparison between the results with and without Coulomb effect. This point is somewhat similar to the case of Wigner delay. Therefore, strictly speaking, the absolute value of response time can not be measured experimentally. Theoretically, the response time can be obtained by comparing the simulation results including long-range Coulomb potential with  short-range ones, but the absolute value of the response time is still uncertain. In TDSE simulations, the response time is analyzed by the instantaneous ionization probability, which is evaluated by the whole populations of continuum states at the instant $t$. But there is uncertainty about how to define the continuum state in a time-dependent system. The results of the MSFA including Coulomb effect depend on the form of Coulomb potential used in the simulation (this dependence partly arises from the classical description of the motion of electrons in the external field and the Coulomb field), so there are also uncertainties. However, the MSFA simulation including Coulomb effect indeed reveals that the response time has the classical correspondence,
which is characterized by the time difference between the instants of electrons exiting the barrier and having instantaneous energy larger than zero,
so it has the physical property of observable time since time in classical mechanics has a clear definition and significance. This is different from the Wigner delay. As mentioned earlier, the Wigner delay itself corresponds to the steady-state problem, in which the state of the system in space (described by the space wave function) is independent of time. The attosecond streaking experiment of the Wigner delay involves bound-continuum transition, and there is no classical correspondence for quantum transition. The tunneling time is related to the motion of electrons in the barrier formed by Coulomb potential and external field, while the tunneling itself has no classical correspondence.

\emph{Relative response time}.-The relative value of Coulomb-induced ionization time lag (strong-field response time) is also of great significance. For example, under the same laser conditions, the difference of the response time of different atoms and molecules reflects the intrinsic difference of the Coulomb potential of the systems, or the difference of the response time of polar molecules in the first and the second half cycles of one laser cycle reflects the intrinsic difference of the sub-cycle dynamics of the system. Therefore, the relative value of the response time can be used to identify the small differences in the internal structure of atoms and molecules at the atomic scale, as well as the subtle differences in the ultrafast dynamics of electrons in the sub-cycle time scale. Although the absolute value of strong-field response time has larger uncertainty, the uncertainty of its relative value is limited in a small range. Therefore, we propose to use a simple and convenient Coulomb-corrected attoclock \cite{Che2} to read the absolute value of the response time corresponding to the most probable trajectory, and then evaluate the relative value of the response time.

\section{Extended discussions}
\emph{Universality}.-The strong-field response time discussed here exists in the interaction of strong laser fields with gases of atoms or molecules. However, we believe that the existence of the response time is general in the interaction of light and matter, such as the interaction between laser and solid \cite{Norreys,Linde,Hohenleutner,Ghimire}, laser and liquid \cite{Heissler,Luu,Zeng}, laser and cluster \cite{Donnelly,Fennel}, laser and plasma \cite{Hamster,Teubner}, etc. For these cases, we anticipate that when the laser field is strong enough and the interaction can be described semiclassically,  the response time and its important effects on the dynamics of the studied system will be easy to identify. When the laser field is weak and the interaction has to be described quantum mechanically, the situation reverses. As discussed above, time is difficult to define in quantum mechanics due to the absence of a quantum time operator. However, the definition of time is unambiguous in classical mechanics.

\emph{Strong-field four-step model}.-The existence of the strong-field response time also strongly suggests  the use of a four-step model including tunneling, response, propagation and rescattering, instead of the well-known three-step one \cite{Schafer,Yang,Corkum,Lewenstein}, to describe strong-field induced ultrafast physical processes of atoms and molecules such as ATI, HHG and NSDI. It allows a great deal of electrons which exit the barrier at the rising part of the laser field also to contribute to the rescattering event \cite{Xie}, with remarkably changing the characteristics of observables such as PMD and HHG spectrum. Such trajectories are prohibited in the general three-step model.

\section{Conclusion}

In summary, 1) the Coulomb-induced ionization time lag (about 100 attoseconds) is not fully considered in the current strong-field experiments of attosecond measurement. The amount of this lag depends on the form of Coulomb potential of atoms and molecules, and  has an important effect on the accuracy of the current  schemes of attosecond measurement. 2) This lag reflects the response time of the electron wave function to the ionization event in strong-laser-matter interactions. The response time is general and has a profound influence on the dynamics of atoms and molecules in the intense laser field. The understanding and quantitative description of the response time need continuous theoretical and experimental efforts. 3) Molecules, especially polar molecules, have complex effects in the external field, such as orientation, alignment, permanent dipole and vibration. These effects may have an important influence on the response time itself or on the observable characteristic quantities characterizing the response time, such as the offset angle of PMD in the elliptical laser field. To understand and probe the response time of electrons in molecules to strong-laser-induced ionization events, it is necessary to develop appropriate strong-field theoretical models and measurement schemes including these effects. 4) Attosecond measurement involves the basic problem of quantum mechanics, which takes time as observation. The basic theory is not perfect, and many concepts are still controversial. Attosecond measurement needs to answer such questions as ``what attosecond-scale process can be measured, what attosecond-scale process can be controlled, and the means and target of attosecond measurement and control".

\emph{Research prospects}.-In the future research on the response time of atoms and molecules in the intense laser field, through  solving TDSE numerically, extracting the real-time information allowed by the numerical experiments, and taking the time-domain analysis of system dynamics as the key breakthrough point, one can deeply study the properties and characterization of response time and its extensive influence on attosecond measurement. By developing the strong-field theory model including the unique effects of molecules, one can study the influences of these effects on the response time, and explore the ultrafast measurement scheme suitable for molecules. On this basis, with the help of the special platform of polar molecules, one can deeply discuss some basic concepts and theoretical problems of attosecond measurement, such as the internal relationship between attosecond measurement and energy-time uncertainty principle, the correspondence between attosecond measurement and classical measurement, the nature of measurable process, and the nature and characterization of response time, etc. These studies will enhance people's understanding of the role of time in quantum mechanics, clarify some controversial conceptual issues in attosecond measurement, and vigorously promote the research of attosecond probing of atoms and molecules.

\emph{Acknowledgments}.-This work was supported by the National Natural Science Foundation of China (Grant No. 91750111), and the National Key Research and Development Program of China (Grant No. 2018YFB0504400).

\end{document}